\documentclass[english,british]{article}
\usepackage[T1]{fontenc}
\usepackage[latin9]{inputenc}
\usepackage{wrapfig}
\usepackage{graphicx}
\usepackage{amssymb}

\makeatletter

\usepackage{wrapfig}

\date{}

\makeatother

\usepackage{babel}

\begin{document}

\title{A Quantum-based Model for Interactive Information Retrieval (extended
version)}

\author{Benjamin Piwowarski\\
University of Glasgow\\
benjamin@bpiwowar.net \and Mounia Lalmas\\
University of Glasgow\\
mounia@dcs.gla.ac.uk}
\maketitle
\begin{abstract}
Even the best information retrieval model cannot always identify the
most useful answers to a user query. This is in particular the case
with web search systems, where it is known that users tend to minimise
their effort to access relevant information. It is, however, believed
that the interaction between users and a retrieval system, such as
a web search engine, can be exploited to provide better answers to
users. Interactive Information Retrieval (IR) systems, in which users
access information through a series of interactions with the search
system, are concerned with building models for IR, where interaction
plays a central role. There are many possible interactions between
a user and a search system, ranging from query (re)formulation to
relevance feedback. However, capturing them within a single framework
is difficult and previously proposed approaches have mostly focused
on relevance feedback. In this paper, we propose a general framework
for interactive IR that is able to capture the full interaction process
in a principled way. Our approach relies upon a generalisation of
the probability framework of quantum physics, whose strong geometric
component can be a key towards a successful interactive IR model.

\global\long\def\dproj#1#2{#1\vartriangleright#2}
\global\long\def\Projector{\mathbb{P}}

\end{abstract}
\global\long\def\bra#1{\left\langle #1\right|}
\global\long\def\ket#1{\left|#1\right\rangle }
\global\long\def\braAket#1#2#3{\left\langle #1\middle|#2\middle|#3\right\rangle }

\global\long\def\braket#1#2{\left\langle #1\middle|#2\right\rangle }
\global\long\def\pr{\mbox{Pr}}
\global\long\def\event#1{\mathbf{#1}}
\global\long\def\kbasis#1{\ket{#1}}
\global\long\def\bbasis#1{\bra{#1}}

\global\long\def\norm#1{\left\Vert #1\right\Vert }

\section{INTRODUCTION}

In less than twenty years, search engines on the Web have revolutionised
the way people search for information. The speed with which one can
obtain an answer to a keyword-based query on the Web is fostering
interaction between search engines and their users. Helping users
to reach relevant material faster will most likely make use of such
rich interaction. Another key to future search systems is the context
that further defines the search, whether it be external~(e.g. time
of the day, location) or internal~(e.g. the interests of the user).

Putting aside the problem of evaluating such contextual and interactive
search~(see~\cite{Robertson2008On-the-history-of-evaluation} for
a discussion on this topic), building models able to explicitly take
into account both is of importance, especially since Information Retrieval
(IR) models seem to have reached maturity and there is an obvious
need to go beyond current state-of-the-art~\cite{Sparck-Jones2006Whats-the-value}. 

There are many reasons why we cannot assume that users will provide
enough information to state an unambiguous Information Need~(IN),
such as a TREC topic description. First, users do not always know
how to express their IN and they sometimes have only a vague knowledge
of what they are looking for. Second, users knowledge and interests
might evolve during the search, thereby modifying their IN. Therefore,
it is important that implicit contextual and interaction {}``information''
become integrated directly into IR models \emph{and} experiments~\cite{Ingwersen2005The-Turn-Integration}.

Beside standard relevance feedback models like the Rocchio algorithm~\cite{Rocchio1971Relevance-feedback}
or the Okapi model~\cite{Walker1999Okapi/Keenbow-at-TREC-8}, some
recent works have attempted to capture context~\cite{Melucci2008A-basis-for-information}
or interaction~\cite{Shen2005Implicit-user}. However, there is not
yet a principled framework that combines both, and that, equally importantly,
tries to capture the different forms of possible interactions, namely,
query (re)formulation, clicks, navigation. Those tasks are all performed
frequently in web searches.

In this paper, we present a framework for interactive and contextual
IR. We view search as a process with two different dynamics: (P1)
The system tries to capture the user IN while (P2) the user cognitive
state, and hence the user IN, is evolving and changing~\cite{Xu2007The-dynamics-of-interactive}.
While the former could be modeled by standard probabilistic models,
we claim that the latter can be better modeled by the generalisation
of probability theory that has been developed in quantum physics~(Section~\ref{sub:measurement}).
Moreover, the strong geometric component of the quantum probability
framework is particularly important since standard IR models rely
on vector spaces and on (some variants of) the cosine similarity~\cite{Rijsbergen2004The-Geometry-of-Information}.
We show how the quantum formalism generalises these latter models~(Section~\ref{sub:superposition}).
In particular, we believe that one strength of the geometric models
in IR is that they are intuitive. Adding a probabilistic view on this
geometry opens the door for new and potentially more powerful IR models. 

Our contribution is to describe how the quantum probability formalism
could be used to build an interactive IR framework, first motivating
the framework in Section~\ref{sec:information-need-space} and then
describing it more in details in Section~\ref{sec:quantum-view}.
We shortly discuss how a working model could be built in Section~\ref{sub:mapping}.
Finally, we detail a direct application of our framework as a principled
generalisation of the Rocchio's algorithm in Section~\ref{sub:mixture}.

\section{Related works}

\label{sec:related-works}

Interactive IR has been actively studied by the Information Science
community~\cite{Ingwersen2005The-Turn-Integration}, and it has been
recognised that there is still little progress regarding the development,
implementation and validation of a model for interactive IR. In particular,
a model that allows for the evolution of the user and the system states
is yet to be developed.

An aspect of interactive IR that has been studied for a long time
now is relevance feedback, where the information about the relevance
of documents is exploited in order to propose a new ordering of the
documents. One of the most well-known algorithm is the Rocchio algorithm~\cite{Rocchio1971Relevance-feedback}
where the query representation in a term vector space is modified
through user interaction. Each time the user deems a document relevant,
the query representation drifts slightly toward the relevant document
representation. Within a more principled probabilistic framework,
the Okapi model can straightforwardly integrate this kind of information
into the statistics used to estimate the model probabilities~\cite{Walker1999Okapi/Keenbow-at-TREC-8}.
In contrast to Rocchio, with Okapi the last feedback has as much importance
as the first ones, which is not a good property for interactive IR
models.

One limitation of these two IR models is that they rely on explicit
relevance feedback, which might not be available or precise enough~\cite{Magennis1997The-potential-and-actual}.
Fortunately, it is possible to adapt these approaches when feedback
is implicit. Shen et al.~\cite{Shen2005Implicit-user} uses a term
vector space representation of a user. The vector is updated when
the user clicks on search result links, based on the summaries of
the documents~(and not on the content of the documents). However,
the update still relies on the Rocchio algorithm which lacks a principled
motivation. In this paper, we show how our framework can extend Rocchio's
algorithm, providing a principled way to model the IN drift.

More general and holistic models have been proposed to build an interactive
retrieval system (e.g. Fuhr~\cite{Fuhr2008A-Probability-Ranking}
and Shen et al.~\cite{Shen2005Implicit-user}) that rely on a decision-theoretic
framework to determine what is the best next action the system should
perform, i.e. what documents should be presented to the user. In such
approaches, decisions are made based on the relevance of documents
when considering past interactions. In this paper, we focus on the
latter problem and do not discuss how to select the best next action
the system has to perform.

Finally, some works have already explored the use of the quantum theory
framework in IR. Firstly, the proposition of using quantum theory
to model IR processes and the interpretation of~(among others) the
standard vector model within the quantum probability formalism have
been detailed in~\cite{Rijsbergen2004The-Geometry-of-Information}.
Following, some works have tried to exploit the formalism, from a
very abstract and generic level~\cite{Arafat2007Foundations-research}
to the formalisation of existing IR models~\cite{Zuccon2008A-Formalization-of-Logical}.
Another work looked at how to represent documents in a space different
from the standard term space~\cite{Huertas-Rosero2009Eraser-Lattices}.
Our contribution with respect to these works is to define a new space,
the IN space, and to clearly map different aspects of user-system
interaction to operations within this space~(Section~\ref{sec:information-need-space}).
In our representation, we also make use of the distinction between
mixtures and superposition, two related but different concepts in
quantum theory~(Section~\ref{sub:superposition}), which proves
to be useful in our representation.

The most related work in that field is that of Melucci's~\cite{Melucci2008A-basis-for-information},
which computes the probability of having a given context $\mbox{\ensuremath{\pr}}_{d}(C)$,
where $\mbox{\ensuremath{\pr}}_{d}$ is the probability distribution
generated by the document vector $d$, and the subspace $C$ is equalled
to the context and is built through user interaction. More specifically,
given a set of documents deemed relevant, either using user feedback
or pseudo-relevance feedback, one can compute a subspace $C$ corresponding
to the principal components of a subspace spanned by those documents.
A document vector fully included in this space will be fully relevant~(probability
of 1), an orthogonal one will be fully irrelevant (zero probability).
Melucci's approach is dual to ours, in the sense that instead of representing
users in an IN space, he considers documents in a contextual space.
Our approach, which relies on an IN space, facilitates the possibility
of using the different quantum evolution mechanisms to model the interaction
between the user and the retrieval system. 

In this paper, we explain how our formalism can exploit implicit~(i.e.
clicks) and explicit feedback, like relevance judgements or query
reformulation. Other forms of interaction like e.g. going back to
the retrieval list, exploration of the result list, etc. could also
be captured by our formalism using the same represent as for e.g.
query (re)formulation, but we do not deal with those in this paper.
To the best of our knowledge, there is no proper theoretical framework
that allows to capture this different feedback information in a principled
way and within a uniform framework.

\section{An Information Need Space}

\label{sec:information-need-space}

We build upon the work described in~\cite{Piwowarski2009Structured-Information}
where a high level description of the framework was first given. Our
working hypothesis is that a \emph{pure}, in the sense that we know
exactly what the user is looking for, user IN can be represented as
a system in quantum physics, i.e. as a unit vector in a Hilbert space%
\footnote{In brief, an inner product vector space defined over the complex field,
see~\cite{Rijsbergen2004The-Geometry-of-Information} for a formal
definition%
}, and that this state evolves while the user is interacting with the
system. 

According to the quantum probability formalism, this (IN) vector generates
a probability distribution over the different subspaces of the Hilbert
space. We make the hypothesis that among other possible uses, such
subspaces can be related to the relevance of documents, therefore
enabling the computation of a relevance score for a document, and
to user interactions~(like typing a query or clicking on a document),
making it possible to exploit them.

From a geometric perspective, using subspaces to describe {}``regions''
of INs has been (sometimes implicitly) studied and motivated in some
works relying on a vector space representation. First, Wang et al.~\cite{Wang2008A-study-of-methods}
studied the problem of negative feedback, finding that this could
be exploited by describing the IN by a set of vectors~(instead of
one as in Rocchio's algorithm). Second, Widdows~\cite{Widdows2003Orthogonal-negation}
has shown that term negation in a query was better modelled with orthogonality,
which implicitly implies the use of subspaces to represent negations
of INs. Finally, Zuccon~\cite{Zuccon2009Semantic-Spaces} has shown
that the cluster hypothesis still holds when representing documents
as subspaces.

Using those IN {}``regions'', the search process would be modelled
as follows. At the very beginning of the search process, the user
IN is underspecified and is a mixture of \emph{all} possible pure
INs. That is, without any information about the user, we can only
know that the user is in one of all the possible IN states with a
probability that depends e.g. on how popular this IN is. 

We believe that using an IN space can model interactive IR since users
change their point of view during a search, and relevance, contrarily
to topicality, is expected to evolve within a search session~\cite{Xu2007The-dynamics-of-interactive}.
More specifically, we can identify two different types of dynamics
within the search process: (P1) The IN becomes increasingly specific
\emph{from a system point of view}, e.g. when a user types some keywords
or clicks on some documents, i.e. the uncertainty is reduced; and
(P2) The IN changes \emph{from a user point of view}. The IN can become
more specific as the user reads some documents, or it can slightly
drift as user interests do. This type of drift has been observed and
analysed by~\cite{Xie2000Shifts-of-interactive} within search sessions.

An example of such a two-part process is described in~\cite{Xu2008Novelty-and-topicality}
where the hypotheses are that a user has a topicality and a novelty
profile, the former providing the context of the IN while the latter
changes during a search session. Our model captures both processes
within the same space, which we refer to as an \emph{Information Need
Space} in this paper.

Whereas the first process can be easily described within a standard
probabilistic framework~(we restrict the IN to subspaces of the whole
space), the latter would benefit from a quantum probability formalism
as the INs can drift from two overlapping subspaces. For example,
consider Figure~\ref{fig:Three-2D} and assume that the possible
user INs lie within a three-dimensional space. If we first restrict
the IN to subspace $A$, then this corresponds to process~(P1): Vectors
orthogonal to $A$ would be discarded. Later, if we know based on
further evidence that the IN subspace is now $B$, then this corresponds
to process~(P2): As vectors belong to $A$, they are not orthogonal
to $B$ and thus projecting them in $B$ makes them drift away from
their original representation. We posit that the classical probabilistic
framework would address the uncertainty of the system view over the
retrieval process~(P1) whereas the quantum probability framework
addresses the changes of the user internal state~(P2). As the quantum
probability framework is a generalisation of the probabilistic one,
we can use the same representation and evolution operators to model
both processes. We describe this framework in the following section.

\section{A quantum view}

\label{sec:quantum-view}

In this section, we describe our framework while also introducing
the framework of quantum probability.

\subsection{Link with standard probability theory}

Quantum probability can be thought of as an extension of classical
probability theory, and relies on linear algebra in Hilbert spaces.
The equivalent of a logical proposition or event is a subspace. As
there is a one-to-one relationship between subspaces and projector
on subspaces, we will switch between one interpretation and the other,
depending on what we want to illustrate. A projector is a special
case of an \emph{observable, }more precisely a yes/no observable:
The observable is binary because a state~(vector) can belong or not
to the associated subspace. It is possible to naturally extend standard
probability theory, by defining a probability density operator over
those subspaces, i.e. by associating with each subspace/projector
a probability.

Next, we first present how to reconstruct a classical probability
space using a Hilbert space, in order to show how the quantum formalism
is a general case of probability theory. We then show how this can
be extended to the quantum probability formalism, and highlight their
differences and uses in our interactive IR framework. 

\begin{wrapfigure}{I}{0.3\columnwidth}%
\selectlanguage{english}%
\centering{}\includegraphics[width=0.3\textwidth]{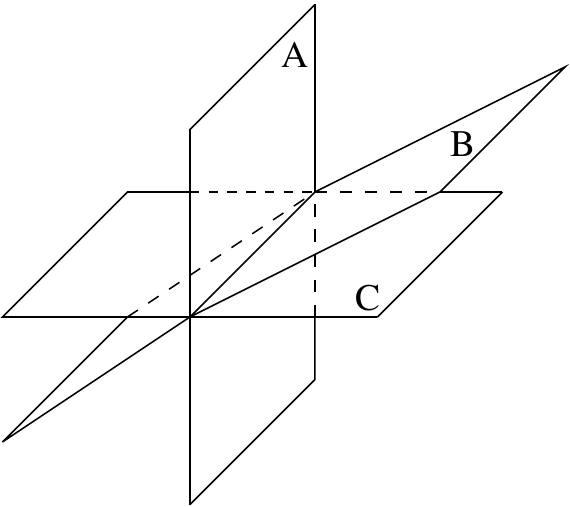}\caption{\selectlanguage{british}%
\label{fig:Three-2D}Three two-dimensional subspaces~(A, B, C) in
a three dimensional space.\selectlanguage{english}
}
\selectlanguage{british}
\end{wrapfigure}%

For simplicity, let us consider a probability space with a finite
sample space $\Omega=\left\{ \omega_{1},\ldots,\omega_{N}\right\} $
of cardinality $N$ where each $\omega_{i}$ corresponds to a \emph{pure}
IN. Moreover, as each $\omega_{i}$ corresponds to a distinct atomic
event, we require that those INs be orthogonal~(i.e. a user with
the IN $\omega_{i}$ will not be interested by a piece of information
answering $\omega_{j}$ where $i\not=j$), a restriction that can
easily be removed within the quantum formalism as discussed in Section~\ref{sub:superposition}.
As usual, a probability distribution is associated with this sample
space, i.e. a probability $\pr\left(\omega_{i}\right)$ is associated
with each \emph{atomic} event in $\Omega$. Then, an event is a subset
$\event A$ of the sample space $\Omega$ whose probability is $\pr\left(\event A\right)=\sum_{\omega\in\event A}\pr\left(\omega\right)$.

In the sequel, we use the bra-ket Dirac notation as done in quantum
theory. Restricted to finite vectorial spaces, a vector of the Hilbert
space is denoted by $\ket x$ while its conjugate transpose is denoted
$\bra x\dot{=}\ket x^{\dagger}$. 

Let us see how our probability space is formalised within the quantum
probability formalism. We associate our sample space $\Omega$ with
one basis of a Hilbert space of dimension $N$. Let us choose an arbitrary
orthonormal basis $\mathcal{B}=\left\{ \kbasis{\omega}\,|\,\omega\in\Omega\right\} $
where each element of the basis $\kbasis{\omega}$ corresponds to
the atomic event $\omega$. These vectors are pairwise orthogonal
and of norm 1. To each atomic event of the sample space is associated
a one-dimensional projector $\Projector_{\omega}$ corresponding to
the one-dimensional subspace defined by $\kbasis{\omega}$, and which
is equal to $\ket{\omega}\bra{\omega}$. Note that it corresponds
to a $n\times n$ matrix where $n$ is the dimension of the Hilbert
space. The probability distribution over the Hilbert space is defined
by a so-called \emph{density operator} $\rho$ defined by 

\begin{equation}
\rho=\sum_{\omega\in\Omega}\pr\left(\omega\right)\Projector_{\omega}\label{eq:mixture}\end{equation}
 where the $\pr\left(\omega\right)$ sum up to 1. Note that a special
case of a density operator is simply $\Projector_{\omega}$ for a
given atomic event $\omega$ when the exact state of the system is
known, e.g. the IN of the user is completely specified. In this case,
the whole probability mass is concentrated onto a single atomic event,
i.e. $\pr\left(\omega\right)=1$.

The event $\mathbf{A}$ is associated with a yes/no observable $O_{\event A}$.
It is a projector associated with the subspace spanned by the basis
vectors $\kbasis{\omega}$ for $\omega\in\event A$. This projector
can be written $O_{A}=\sum_{\omega\in\event A}\Projector_{\omega}$.
In the following formula where $\mbox{tr}$ denotes the trace operator,
we give both quantum and standard definitions to show how they relate:
\begin{equation}
\underbrace{\vphantom{\sum_{w\in\event A}}\pr_{\rho}\left(O_{A}\right)\dot{=}\mbox{tr}\left(\rho O_{A}\right)}_{\mbox{quantum definition}}=\mbox{tr}\left(\sum_{\omega\in\Omega}\pr\left(\omega\right)\Projector_{\omega}\sum_{\omega\in\event A}\Projector_{\omega}\right)=\underbrace{\sum_{w\in\event A}\pr\left(\omega\right)\dot{=}\pr(\mathbf{A})}_{\mbox{standard definition}}\label{eq:trace-pr}\end{equation}
The equality is due to the fact that $\Projector_{\omega}\Projector_{\omega^{\prime}}=0$
whenever $\omega\not=\omega^{\prime}$, as we assumed the vectors
were pairwise orthogonal. This is exactly the way of computing a probability
by decomposing it into atomic and incompatible events~(as shown on
the right side of the equation). Without giving too many details,
we point out that the quantum definition is more general because $O_{A}$
can be expressed in a different basis than $\mathcal{B}$.

We have just defined a formalism that allows us to express a classical
probability distribution using linear algebra. All the information
about the probability distribution is contained into a \emph{density
operator} $\rho$, and it can be shown that for \emph{any} probability
distribution over a Hilbert space there exists a corresponding density
operator~\cite[p. 81]{Rijsbergen2004The-Geometry-of-Information}.
From a practical point of view, the above description of standard
probabilities with Hilbert spaces unlocks the potential of defining
probabilities through geometric relationships, and permits a generalisation
to a non standard probability formalism, which we describe in the
next section.

We posit that at this level, we are able to model the first component
of the search process, which corresponds to finding the right \emph{subspace}
of the IN, i.e. in classical terms to find the subset of the IN sample
space. However, it is intuitive to think that INs are not mutually
exclusive. We make the hypothesis that such a non-exclusiveness is
captured by the geometry of IN space, and this can be modelled within
a quantum probability formalism, as discussed in the next section.

\subsection{Superposition, mixtures and information needs}

\label{sub:superposition}

Now we turn to how probability theory is generalised. We introduce
the notion of superposition and mixture, and relate them to their
use in our model of interactive IR. Said shortly, superposition relates
to an ontologic uncertainty~(the system state is perfectly known,
but some events are true \emph{only} with a given probability) whereas
mixture relates to standard probabilistic uncertainty~(the system
is in one of the states with a given probability). Superposition is
a salient characteristic of quantum probabilities and is important
since it gives us a way to represent geometrically new INs while the
quantum probability framework ensures we can still compute probabilities
for the new INs. Mixture and superposition gives us more flexibility
in the way we can represent our current state of knowledge of an IN.

Let us illustrate this with an example. Suppose that $\ket{\omega_{T}}=\left(\begin{array}{cc}
1 & 0\end{array}\right)^{\dagger}\mbox{ and }\ket{\omega_{L}}=\left(\begin{array}{cc}
0 & 1\end{array}\right)^{\dagger}$ form a basis of the IN space. Suppose the former represents the IN
of a user looking for information about tigers~(T) and the latter
about lions~(L). In order to represent a user looking for a tigron~(the
offspring of a tiger and a lion), we assume that this can be represented
by the vector $\ket{\omega_{TL}}=\frac{1}{\sqrt{2}}\left(\ket{\omega_{T}}+\ket{\omega_{L}}\right)$
which is a \emph{superposition} of two INs, where the $\frac{1}{\sqrt{2}}$
factor ensures $\ket{\omega_{TL}}$ norm is one. This is a strong
assumption which we will study when experimenting with the framework.
Aerts and Gabora~\cite{Aerts2005A-Theory-of-Concepts-II} worked
on how to combine concepts in a (quantum) vector space, but use spaces
of increasing dimensionality to do so~(through the use of a tensor
product). As a final remark on superposition of INs, we would like
to note that complex numbers could be used to combine INs, e.g. to
distinguish tigrons~(the tiger is the father) from ligers~(the lion
is the father), and that superposition is not restricted to topicality.
For instance, assuming that we know how to represent a user searching
for a paragraph and a user searching for a chapter, we could imagine
representing a user looking for a paragraph as a superposition of
both.

\global\long\def\TorL{T\vee L}
The superposed IN $\ket{\omega_{TL}}$ is quite different to a user
who is equally interested by tigers or lions, which would be represented
as a \emph{mixture} of the INs $\ket{\omega_{T}}$ and $\ket{\omega_{L}}$.
Formally, this IN would be associated with a density operator$\rho_{T\vee L}=\frac{1}{2}\left(\rho_{T}+\rho_{L}\right)$
where $\rho_{L}$ and $\rho_{T}$ are respectively the projectors
associated with $\ket{\omega_{T}}$ and $\ket{\omega_{L}}$. For example,
$\rho_{T}=\ket{\omega_{T}}\bra{\omega_{T}}$. The density operator
$\rho_{\TorL}$ is to be interpreted by saying that with probability
one half the IN is about tigers~(or equivalently about lions).

We can see also the difference if we represent the densities by their
matrices in the $\left(\ket{\omega_{T}},\ket{\omega_{L}}\right)$
basis. We have the mixture of IN $\rho_{\TorL}=\frac{1}{2}\left(\begin{array}{cc}
1 & 0\\
0 & 1\end{array}\right)$ which is different from the pure IN $\rho_{TL}=\frac{1}{2}\left(\begin{array}{cc}
1 & 1\\
1 & 1\end{array}\right)$. An important observation is that these different densities imply
different probabilities. Let us suppose that the relevance of a document
corresponds to a yes/no observable, and that the relevance of a document
about lions~(respectively tigers, tigrons) are represented by the
projectors~(yes/no observables) $O_{L}$, $O_{T}$ and $O_{TL}$
associated with the subspaces generated by $\ket{\omega_{T}}$, $\ket{\omega_{L}}$
and $\ket{\omega_{TL}}$, respectively. For example, $O_{T}=\ket{\omega_{T}}\bra{\omega_{T}}$.
According to Eq.~(\ref{eq:trace-pr}), we can compute the probability
of relevance of the different documents, which gives:

\[
\begin{array}{cccc}
\pr_{\rho_{TL}}\left(O_{L}\right)= & \mbox{\ensuremath{\pr_{\rho_{\TorL}}\left(O_{L}\right)}=\ensuremath{\frac{1}{2}}\mbox{ and }} & \pr_{\rho_{TL}}\left(O_{TL}\right)=1\not= & \pr_{\rho_{\TorL}}\left(O_{TL}\right)=\frac{1}{2}\end{array}\]

Interestingly, we cannot distinguish the probability of relevance
of the document about lions when the IN is about either tigers and
lions or about tigrons~(two first probabilities) but there are two
reasons for this: In the former, the probability $\frac{1}{2}$ is
caused by the discrepancy between the IN and the document, whereas
in the second case the probability is due to the fact that the document
only covers a part of the information need. Next, thanks to the quantum
formalism the probabilities for the same INs are different when we
evaluate the relevance of the document about tigrons~(two last probabilities).
We thus benefit from a two-dimensional space to distinguish different
INs that would be expressed similarly in a standard vector space model.
One consequence is that if we search for a set of documents that satisfy
$T$ or $L$, we would have two different types of documents (about
tigers and lions, assuming each document covers one IN only) whereas
one document would satisfy $TL$. 

Mixtures are also useful to represent the IN density operator $\rho_{0}$
at the very beginning of the information retrieval process, as we
do not know which state the user is in. We would define the initial
IN density operator as $\rho_{0}=\sum_{i}\pr_{\mbox{i}}\Projector_{\mbox{i}}$
where $i$ ranges over all the possible information needs and $\pr_{i}$
is the probability that a random user would have the IN $i$ when
starting a search. Using the mixture is also motivated by the fact
that we deal with classical undeterminism, i.e. we know the user is
in a given state but we do not know which. The mixture can also be
thought as a set of vectors describing all the possible INs, each
vector being associated with a probability. This representation is
particularly useful in the next section where we show how this initial
IN $\rho_{0}$ is transformed through interactions.

\subsection{Measurement and Interaction }

\label{sub:measurement}\global\long\def\obs#1{O_{#1}}

Beside differentiating mixture and superpositions, the quantum formalism
has also consequences for computing a conditional probability. These
consequences are linked to the way a measurement is performed in quantum
physics. We use measurement to model interaction and describe in this
section both how the measurement modifies the density operator$\rho$
and how we link measurement to the different interactions.

For simplicity, we now use $A$ to denote the related yes/no observable,
subspace or projector. Since there is a one-to-one correspondence
between them~\cite{Rijsbergen2004The-Geometry-of-Information}, they
can be used to denote the same thing albeit in a different context.
Given a system density operator $\rho$, if we observe $O_{A}$, the
new density operator denoted $\rho\vartriangleright O_{A}$ is defined
by \begin{equation}
\dproj{\rho}{O_{A}}=O_{A}\rho O_{A}/\mbox{tr}\left(\rho O_{A}\right)\label{eq:conditioning}\end{equation}
This amounts to restricting $\rho$ to the subspace defined by $O_{A}$
and ensuring that $\dproj{\rho}{O_{A}}$ is still a density operator.
The effect of the restriction is to project every IN of the mixture
$\rho$ onto the subspace defined by $\obs A$~(with some renormalisation
to ensure the probabilities still sum up to 1). One can readily verify
that the probability of $O_{A}$ with respect to $\rho\vartriangleright O_{A}$
is 1. It means that when $A$ has just be measured, we know it is
true at least until further interaction~(or in general, evolution)
modifies the density operator. Measurement can be thought as a generalisation
of conditionalisation, as we can compute the conditional probability
of $O_{A}$ given $O_{B}$, or more precisely of measuring $O_{A}$
knowing that we have measured $\obs B$, as $\pr_{\rho}\left(\obs B|\obs A\right)=P_{\dproj{\rho}{\obs A}}(\obs B)$. 

In quantum theory, the order of the measurements is important, since
in general the densities $\dproj{\dproj{\rho}{\obs A}}{\obs B}$~(applying
two times the Eq.~(\ref{eq:conditioning}), for $\obs A$ and then
for $\obs B$) and $\dproj{\dproj{\rho}{\obs B}}{\obs A}$ are different.
It is a desirable property whenever subsequent measurements of a system
should yield different results, which is the case in interactive IR:
The sequence of interactions represents the evolution of the user,
and should be taken into account. A user drifting from an IN~(e.g.
hotels in Barcelona) to another~(e.g. museum in Barcelona) is not
the same as the reverse, which illustrates the adequacy of the quantum
formalism to handle such drifts. This is illustrated by Figure~\ref{fig:Three-2D},
where visually it can be seen that measuring $\obs B$~(hotels) then
$\obs C$~(museums) is different from the reverse, since in the first
case the IN vectors will lie in the subspace $C$ whereas they would
lie in $B$ in the other case.

Starting with the initial density operator $\rho_{0}$~(section~\ref{sub:superposition}),
we make the assumption that each implicit or explicit interaction
between the IR system and the user corresponds to a measurement, i.e.
that every interaction is associated with a yes/no observable $O$.
After the interaction, we can recompute the IN density operator using
Eq.~(\ref{eq:conditioning}). For example, a user whose internal
context is associated as $O_{\mbox{user}}$, who asked a query associated
with $O_{q_{1}}$ and deemed a document relevant~(associated with
$O_{d_{1}}$), would be represented by a density operator $\dproj{\rho_{0}}{\dproj{O_{\mbox{user}}}{\dproj{O_{q_{1}}}{O_{d_{1}}}}}$.
Among other users, this density operator can be used to predict the
relevance of other documents.

In the following, we make our framework more concrete: In Section~\ref{sub:mapping}
we show how some interactions would be mapped to an observable. We
argue that direct measurement can be too harsh in some cases and show
how a generalisation of Rocchio's rule can be used to overcome this
problem in Section~\ref{sub:mixture}.

\subsubsection{Mapping interactions to observables}

\label{sub:mapping}

In order to map interactions to observables, we restrict to the topical
relevance and assume a vector space where dimensions are associated
with terms. How to deal with more relevance dimensions is left for
future work. We also assume we know how to compute the initial density
operator $\rho_{0}$~--~which could be approximated using the document
representation described next.

Giving the current IN density operator $\rho_{t}$, we can compute
the probability of relevance $\pr_{\rho_{t}}\left(O_{d}\right)$ of
a document $d$ , provided $O_{d}$ is the observable associated with
the relevance of document $d$. To build such an observable, and as
a first approximation, we can suppose that each paragraph $p$ corresponds
to exactly one IN $\ket{\omega_{p}}$, and hence that its representation
is a one dimensional subspace. It is then possible to compute the
subspace spanned by the vectors $\left\{ \omega_{p}\right\} $ corresponding
to the different paragraphs, and use this subspace to represent the
document relevance. When a user deems a document relevant, we could
use the same representation to update the current IN $\rho_{k}$.
In that case, we would have the new IN density operator $\rho_{t+1}=\dproj{\rho_{t}}{O_{d}}$. 

The first possible type of interaction would be the (re)formulation
of a query by a user. We would associate to a given query a subspace/observable
$O_{q}$, and update the current probability density operator $\rho_{t}$
to $\rho_{t+1}=\dproj{\rho_{t}}{\obs q}$. A representation of the
query could for example be computed through pseudo-relevance feedback
provided we know how to represent the documents: The subspace associated
with $\obs q$ would then be the subspace spanned by the observables
representing the top-ranked documents~(by a standard IR algorithm).
For example, in Figure~\ref{fig:Three-2D}, if $A$ and $B$ correspond
to two different top-ranked documents for a given query, then $\obs q$
would correspond to the whole three dimensional space~(i.e. the join
of subspaces $A$ and $B$). Another way to compute the query observable
$O_{q}$, without relying on an external model, would be the union
of the subspaces representing the paragraphs where each term of the
query appears.

Here, we give one illustration of the usefulness of the quantum formalism
for an interactive IR framework. The query observable $O_{q}$~(or
the document observable $O_{d}$) can be used to detect if a user's
change of mind is too important to be a simple drift, an important
feature an interactive IR system should have~\cite{Xie2000Shifts-of-interactive}.
Within the quantum framework, we use the same geometric representation
to both update the density operator knowing an event and to compute
the probability of this event. Indeed, when at time $t$ the user
types a new query $q^{\prime}$, we can compute the probability of
the query according to the current IN density operator $\rho_{t}$,
i.e. compute $\pr_{\rho_{k}}\left(O_{q^{\prime}}\right)$. Based on
this value, our IR system would decide that the user switched to a
new IN, and react accordingly.

\subsubsection{A probabilistic generalisation of Rocchio's update rule}

\label{sub:mixture}

Sometimes measurement can be too harsh, either because we are not
sure we made the observation~(as in relevance feedback) or because
we do not expect the user IN to completely drift to the INs covered
by e.g. a clicked document. Said otherwise, we want to give some inertia
to the user and this is exactly what Rocchio's algorithm does. We
show how this algorithm can be extended in a principled way with our
formalism.

Consider a user looking for information about tigers or lions, which
can be represented, respectively, as the subspaces $A$ and $C$ of
Figure~\ref{fig:Three-2D}. If the user deems relevant a document
about tigers, the updated density operator would only be made of INs
belonging to the subspace $A$, which is orthogonal to $C$. Documents
about lions would be henceforth considered as non-relevant since $\pr_{\dproj{\rho}{\obs A}}\left(C\right)=0$
whatever the density operator $\rho$ is. In order to give some inertia
to the underlying dynamics, we could as in Rocchio's algorithm use
a parameter $\alpha$ such that the new density operator combines
both the last IN density operator and the new one, i.e. $\rho_{t+1}=\alpha\left(\dproj{\rho_{t}}{\obs A}\right)+\left(1-\alpha\right)\rho_{t}$.
This formula is interpreted as the fact that with probability $\alpha$
the user IN has been projected onto the subspace defined by document
$A$, and with probability $1-\alpha$ it remained the same. It can
be said that the above formula is a principled generalisation of the
Rocchio update rule. However, it is not equivalent to it, since each
IN vector composing the density operator $\rho_{t}$ is (in general,
i.e. when it does not belong to subspace $A$) decomposed into two
vectors, one projected onto $A$ while the other remains the same
whereas with Rocchio, we only deal with one vector, which has limitations
e.g. when using negative feedback~\cite{Wang2008A-study-of-methods}.

\section{Conclusion}

We proposed a new interactive IR framework, which exploits the strong
connection between geometry and probabilities present in the quantum
probability formalism. Our framework allows for a principled and geometric
mapping of user interactions into an IR model. In particular, we show
how to handle click/relevance feedback and query reformulation. How
to use the latter information has not been explored in IR so far,
beside providing query recommendation. Other forms of interaction~(e.g.
navigation) would fit our framework, through the definition of associated
subspaces. Beside measurement, the quantum framework is powerful enough
to provide other types of evolution of the IN density operator. This
would provide a way to predict how a user might evolve, e.g. in order
to predict that users looking for hotels might look for museums in
a town. 

Diversity and novelty are two important components of interactive
IR system~\cite{Xu2008Novelty-and-topicality} that we did not discuss.
We believe that the non-standard logic associated with the subspaces
would be a useful property for such purposes, since we could represent
a set of documents as the union subspace spanned by the different
document subspaces. The relationship between the union subspace and
the possible information needs provide enough information to compute
how the possible INs are addressed and how novel the documents are. 

From a more practical perspective, future work includes experimenting
with different possible ways of mapping interactions to observables,
dealing with complexity issues, in order to build an interactive IR
model based on the ideas we developed in this paper.

\section*{Acknowledgments}

This research was supported by an Engineering and Physical Sciences
Research Council grant (Grant Number EP/F015984/2). Mounia Lalmas
is currently funded by Microsoft Research/Royal Academy of Engineering. 

\bibliographystyle{abbrv}
\bibliography{IQIR}

\end{document}